\documentclass[twoside,english]{elsarticle}
\usepackage[latin9]{inputenc}
\usepackage{geometry}
\usepackage{textcomp}
\usepackage{graphicx}

\makeatletter

\newcommand{\noun}[1]{\textsc{#1}}

\date{\today}
\makeatletter
\def\ps@pprintTitle{%
 \let\@oddhead\@empty
 \let\@evenhead\@empty
 \def\@oddfoot{ \hfil \today}%
 \let\@evenfoot\@empty}
\makeatother

\usepackage[hyphens]{url}
\usepackage{hyperref}
\hypersetup{colorlinks=true}

\makeatother

\usepackage{babel}
\begin{document}

\begin{frontmatter}{}

\title{An extension of the open-source \textit{\noun{porousMultiphaseFoam}}
toolbox dedicated to groundwater flows solving the Richards' equation}

\author{Pierre~Horgue}

\author{Jacques~Franc}

\author{Romain~Guibert}

\author{Gérald~Debenest}

\address{INPT, UPS, IMFT (Institut de Mécanique des Fluides de Toulouse),
Université de Toulouse, Allée Camille Soula, F-31400 Toulouse, France
and CNRS, IMFT, F-31400 Toulouse, France}
\begin{abstract}
In this note, the existing \noun{porousMultiphaseFoam} toolbox, developed
initially for any two-phase flow in porous media is extended to the
specific case of the Richards' equation which neglect the pressure
gradient of the non-wetting phase. This model is typically used for
saturated and unsaturated groundwater flows. A Picard's algorithm
is implemented to linearize and solve the Richards' equation developed
in the pressure head based form. This new solver of the \textit{\noun{porousMultiphaseFoam}}
toolbox is named \emph{groundwaterFoam}. The validation of thesolver
is achieved by a comparison between numerical simulations and results
obtained from the literature. Finally, a parallel efficiency test
is performed on a large unstructured mesh and exhibits a super-linear
behavior as observed for the other solvers of the toolbox. \end{abstract}
\begin{keyword}
Porous medium \sep Unsaturated flow \sep OpenFOAM \sep Richards'
equation \sep Picard's algorithm
\end{keyword}

\end{frontmatter}{}

\section{Introduction}

The modeling and understanding of fluid flow in unsaturated soils
is an important problem in a wide range of scientific domains, such
as environmental engineering or groundwater hydrology. Two-phase
flow in porous media can be modeled by solving the mass conservation
equation for each phase where the phase velocities are expressed using
a generalized Darcy's law \citep{Muskat1949}. However, a classical
approach commonly used in soils science consists in neglecting the
pressure gradient in the non-wetting phase (typically the air) to
reduce the two-phase flow to one equation, the so-called Richards'
equation \citep{Richards1931,Hillel1980}.

Several softwares has been developed to solve the Richards' equation
and some of these developments have already been done using the OpenFOAM
platform \citep{Jasak1996,Rusche2002a}. We can cite the example of
\citet{Liu2012} who developed a saturated-unsaturated groundwater
flow solver based on the Picard's algorithm. This solver includes
several features such as the different forms of the Richards equation
(pressure-based and mixed-form), three convergence criteria and specific
boundary conditions. More recently, another Richards' solver has also
been proposed for the OpenFOAM platform \citep{Orgogozo2014}. Both
initiatives have been shown to have good parallel efficiency.

In a previous work, an open-source toolbox based on OpenFOAM and dedicated
to the simulation of multiphase flow in porous media as been developed
and validated \citep{Horgue2015}. Based on the IMPES method (Implicit
Pressure Explicit Saturation) \citep{J.W.Sheld}, this toolbox includes
the commonly used porous media models (relative permeability, capillary
pressure), specific boundary conditions and validation cases. A good
parallel efficiency has also been demonstrated. This project is still
under development and the toolbox is freely available \citep{Horgue}.

To expand the possibilities and the application fields of the porous
media toolbox, this work proposes to implement a version of the Richards'
equation following the formalism of the toolbox and re-using as much
as possible the existing libraries. First, the mathematical model
and the formulation chosen are presented. In Sec. \ref{sec:Numerical-implementation},
the numerical implementation is developed with the different choices
in terms of time step determination, algorithm, etc. The solver is
then validated and evaluated in terms of parallel efficiency in Sec.
\ref{sec:Validation}. In the following, italic style refers to \textit{solver}s,
small capitals style to \textsc{libraries}, and typewriter style to
\texttt{directories}.

\section{Mathematical model\label{sec:Mathematical-model}}

Three major forms of the unsaturated mass conservation equation exist
in the literature: the pressure head-based, the saturation-based or
the mixed-form formulation. The pressure head-based formulation has
been chosen as this formulation is closed to the previously developed
solvers of the toolbox \citep{Horgue2015}. The Richards' equation
in the pressure head based formulation reads
\begin{equation}
C(h)\frac{\partial h}{\partial t}-\nabla\cdot\left[K_{S}(h)\nabla\left(h+z\right)\right]=0,
\end{equation}

\noindent where $h$ is the pressure head, $C(h)$ the capillary capacity
depending on the head pressure, $K_{S}(h)$ the hydraulic conductivity
and $z$ the elevation. This equation can be formulated as

\begin{equation}
C(h)\frac{\partial h}{\partial t}-\nabla\cdot\left[M_{\theta}\left(\rho_{\theta}\parallel\mathbf{g}\parallel_{2}\nabla h-\rho_{\theta}\mathbf{g}\right)\right]=0,\label{eq:richards}
\end{equation}

\noindent where $\rho_{\theta}$ is the phase density, $\parallel\mathbf{g}\parallel_{2}$
the magnitude of the gravity field and $M_{\theta}$ the phase mobility
of the phase defined as

\begin{equation}
M_{\theta}=\frac{Kk_{r,\theta}}{\mu_{\theta}},
\end{equation}
where $K$ is the intrinsic permeability of the porous medium, $\mu_{\theta}$
the liquid viscosity and $k_{r,\theta}$ the relative permeability.
The saturated hydraulic conductivity $K_{S}$, commonly used for fluid
flow in unsaturated soils, is then directly related to the rock intrinsic
permeability following:
\begin{equation}
K=\frac{\mu_{\theta}K_{S}}{\rho_{\theta}\parallel\mathbf{g}\parallel_{2}}.
\end{equation}

Note that the relative permeability $k_{r,\theta}$ is expressed as
a function of saturation $\theta$ to re-use the already implemented
relative permeability models (\citet{Brooks1964,VanGenuchten1980}).
Using this formulation, only two functions need to be added in the
\noun{capillaryModel} library. The first one allows to compute the
saturation $\theta$ from the pressure head $h$, which gives, for
the Van Genuchten model,

\begin{equation}
\theta(h)=\left\{\begin{array}{ll}
\frac{\theta_{s}-\theta_{r}}{\left((1+(\alpha|h|))^{n}\right)^{m}}+\theta_{r} & \forall \, h<0\\
\theta_{s} & \forall \, h \geq 0
\end{array}\right. \end{equation} 

\noindent where $\theta_{s}$ and $\theta_{r}$ are respectively the
saturated and residual saturations, and $\alpha$ and $m$ the Van
Genuchten's parameters. The second function computes the capillary
capacity
\begin{equation}
C(h)=\frac{\alpha m\left(\theta_{s}-\theta_{r}\right)}{1-m}\left(\theta_{e}\right)^{\frac{1}{m}}\left(1-\left(\theta_{e}\right)^{\frac{1}{m}}\right)^{m}
\end{equation}

\noindent where $\theta_{e}$ is the effective saturation given by

\[
\theta_{e}=\frac{\theta(h)-\theta_{r}}{\theta_{s}-\theta_{r}}.
\]

\noindent The total mobility $M$ is defined as 
\begin{equation}
M=M_{\theta}\rho_{\theta}\parallel\mathbf{g}\parallel_{2}
\end{equation}

\noindent which allows to directly use the existing \noun{darcyGradPressure}
boundary condition for the pressure head field $h$. When using this
boundary conditions, the solver will look up at the fixed value for
the velocity field $U$, and the value of total mobility $M$ to set
the pressure head gradient necessary to impose the fluid velocity.
Readers can refer to the work of \citet{Horgue2015} for more details
about the \noun{darcyGradPressure} boundary condition.

\section{Numerical implementation\label{sec:Numerical-implementation}}

Different iterative techniques can be used to solve the non-linear
problem expressed in Eq. (\ref{eq:richards}) including Picard and
Newton methods. The Picard method has been implemented in this work
as it the simplest and the more robust technique. Note that a better
convergence rate can be obtained with Newton methods but this requires
the computation of a Jacobian matrix (increasing the RAM memory required).

\subsection{Picard's algorithm}

In the Picard method, the pressure-head field $h^{n+1,m+1}$ for the
iteration $m+1$ of the algorithm is computed as:

\begin{equation}
C(h^{n+1,m})\frac{h^{n+1,m+1}-h^{n}}{\Delta t^{n}}-\nabla\cdot\left(\rho_{\theta}\parallel\mathbf{g}\parallel_{2}M_{\theta}^{n+1,m}\nabla h^{n+1,m+1}\right)+\nabla M_{\theta}^{n+1,m}\cdot\rho_{\theta}\mathbf{g}=0\label{eq:richards_discretized}
\end{equation}

\noindent with $h^{n}$ the head pressure value at the last time $n$
and $M^{n+1,m}$the phase mobility computed using the last iteration
$h^{n+1,m}$. The loop occurs until the Picard residual $r_{Picard}$
satisfies:
\begin{equation}
r_{Picard}=max\left(\mid h^{n+1,m+1}-h^{n+1,m}\mid\right)<\epsilon{}_{Picard}
\end{equation}

\noindent where $\epsilon{}_{Picard}$ is the user-defined Picard
tolerance.

\subsection{Time-step \label{sub:Time-step}}

A simple heuristic way has been chosen as proposed in \citep{williams1999evaluation}
for time step determination with a stabilization parameter to avoid
too sharp time-step evolution. This includes three user-defined numbers
of iterations ($n_{maxIter,Picard}$, $n_{minIter,Picard}$ and $n_{maxIter,stabilization}$)
and two time-step factors ($f_{\Delta t,increase}$ and $f_{\Delta t,decrease}$).
After the Picard algorithm has converged using $n_{iter,Picard}$
iterations, three different situations can occur:
\begin{enumerate}
\item $n_{iter,Picard}>n_{maxIter,Picard}$, the current time step is too
large and $\Delta t^{n+1}=f_{\Delta t,decrease}\times\Delta t^{n}$.
\item $n_{minIter,Picard}\leq n_{iter,Picard}\leq n_{maxIter,Picard}$,
the time step remains unchanged $\Delta t^{n+1}=\Delta t^{n}$.
\item $n_{iter,Picard}<n_{minIter,Picard}$:

\begin{enumerate}
\item the stabilized iteration counter is increased: $n_{iter,stabilized}=n_{iter,stabilized}+1$
\item If $n_{iter,stabilized}=n_{maxIter,stabilization}$, then the time
step increases $\Delta t^{n+1}=f_{\Delta t,increase}\times\Delta t^{n}$
and the counter is reseted ($n_{iter,stabilized}=0$).
\end{enumerate}
\end{enumerate}

\subsection{Algorithm}

The global algorithm for each time step consists in:
\begin{enumerate}
\item While $r_{Picard}>\epsilon{}_{Picard}$

\begin{enumerate}
\item solve Richards' equation (\ref{eq:richards_discretized})
\item update flow properties (relative permeabilities, capillary pressure)
\item compute Picard residual $r_{Picard}$
\item if $n_{iter,Picard}>2\times n_{maxIter,Picard}$: break loop, accept
current solution and display warning message
\end{enumerate}
\item Compute $\Delta t$ for the next iteration (see Sec. \ref{sub:Time-step}).
\end{enumerate}

\subsection{Code structure}

The program \emph{groundwaterFoam,} solving the Richards' equation
for an heterogeneous isotropic permeability field ($K$ is an heterogeneous
scalar field) have been added to the \noun{porousMultiphaseFoam} toolbox.
Note that, following the example of \emph{impesFoam} and \emph{anisoImpesFoam},
it is possible to develop a Richards' solver handling anisotropic
permeability fields. The \noun{capillarityModels} functions have been
modified to compute saturation $\theta$ and capillary capacity $C(h)$
from head pressure $h$. Note that the Van Genuchten model is currently
the only model implemented in the toolbox.

Three test cases have been added in the \texttt{groundwaterFoam-tutorials}
folder of the toolbox. The \texttt{1Dinfiltration} simulation is used
to validate the developed solver (see Sec. \ref{sub:Validation-case})
and provides an example of the solver use. The \texttt{1Dinfiltration}\_Ufixed
is close to the previous validation case but using the \noun{darcyGradPressure}
boundary condition (which set the value of the velocity field). The\emph{
}\texttt{realCase} provides an example on a more complex geometry
based on real topographic dataset and has been used to evaluate parallel
efficiency (see Sec. \ref{sub:Parallel-efficiency}).

\section{Numerical simulations\label{sec:Validation}}

\subsection{Validation case\label{sub:Validation-case}}

The vertical 1D water infiltration problem proposed for validation
is derived from the work of \citet{Celia1990} and has been used in
several studies \citep{Rathfelder1994,Kavetski2001}. The column of
New Mexico soils is modeled using the following parameter:
\begin{itemize}
\item $K_{s}=0.00922$ cm.s\textsuperscript{-1} (corresponding to $K=9.4.10^{-12}$
m\textsuperscript{2}),
\item $\theta_{r}=0.102$ and $\theta_{s}=0.368$,
\item $\alpha=0.0335$ cm\textsuperscript{-1},
\item $m=1-\frac{1}{n}=0.5$,
\item $\mu_{\theta}=1\cdot10^{-3}$ Pa.s,
\item $\rho_{\theta}=1\cdot10^{3}$ kg.m\textsuperscript{-3}.
\end{itemize}
The boundary condition on the top of the column is initialized to
$h=-75$ cm (corresponding to $\theta=0.20037$) while the head pressure
is uniformly distributed in the column $h=-1000$ cm (corresponding
to $\theta=0.10994$). The domain is discretized using $200$ computation
cells and the test case is directly available in the toolbox tutorials
(\texttt{1Dinfiltration} folder).

\begin{figure}
\centering{}\includegraphics[angle=-90,origin=c,scale=0.5]{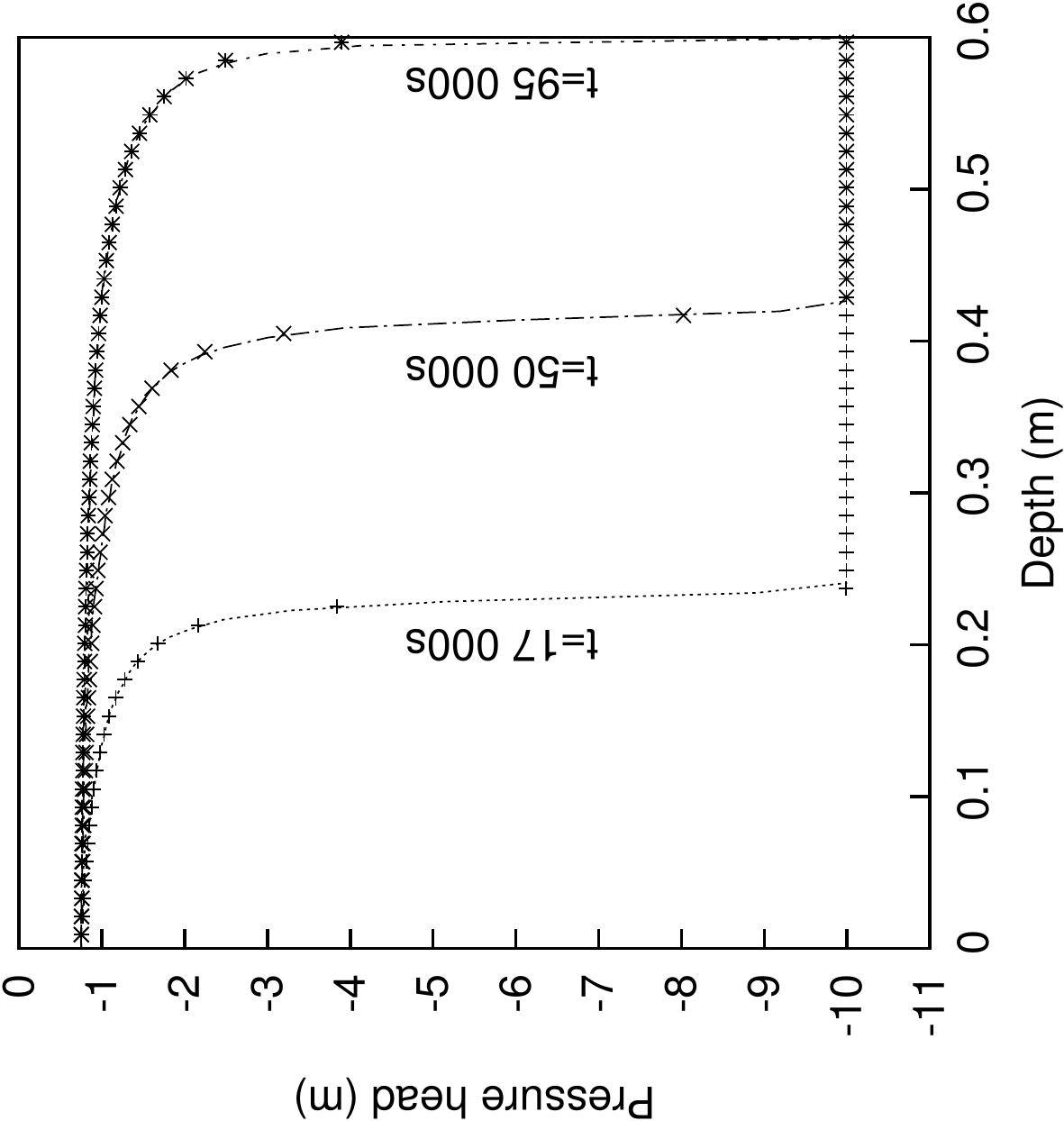}\caption{\label{fig:1D-infiltration-case}Pressure head profiles at various
times for the 1D infiltration case (lines are reference results from
\citet{Kavetski2001}) }
\end{figure}

The comparison between simulations and the reference solution (numerical
results extracted from the work of \citet{Kavetski2001}) presented
in Fig. \ref{fig:1D-infiltration-case} shows a good agreement and
validates the code.

\subsection{Parallel efficiency\label{sub:Parallel-efficiency}}

\begin{figure}
\centering{}\includegraphics[width=10cm]{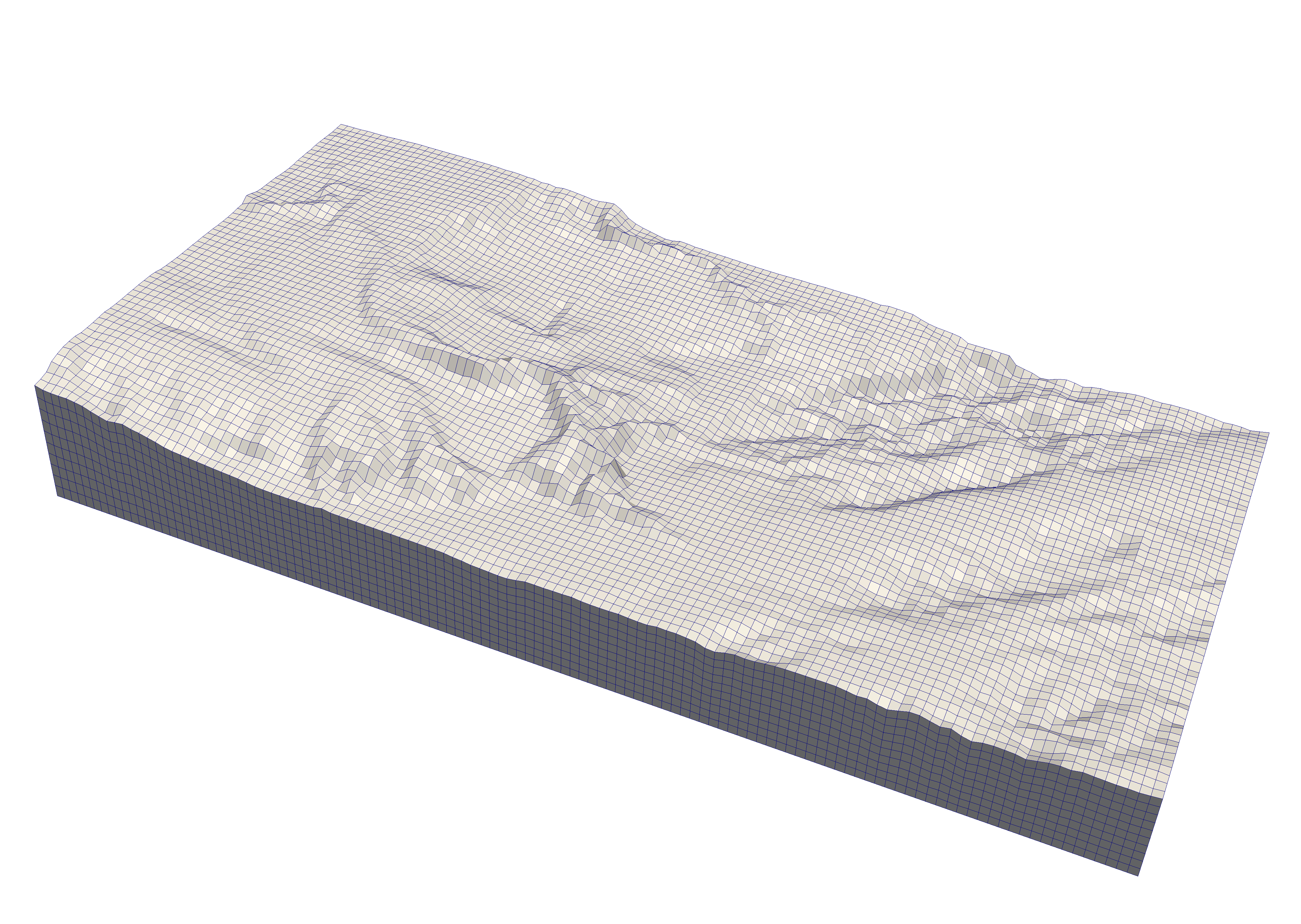}\caption{Unstructured coarse mesh based on real topographic dataset (aspect
ratio of the visualization $1:1:4$)\label{fig:Unstructured-coarse-mesh}}
\end{figure}

\begin{figure}
\centering{}\includegraphics[width=10cm]{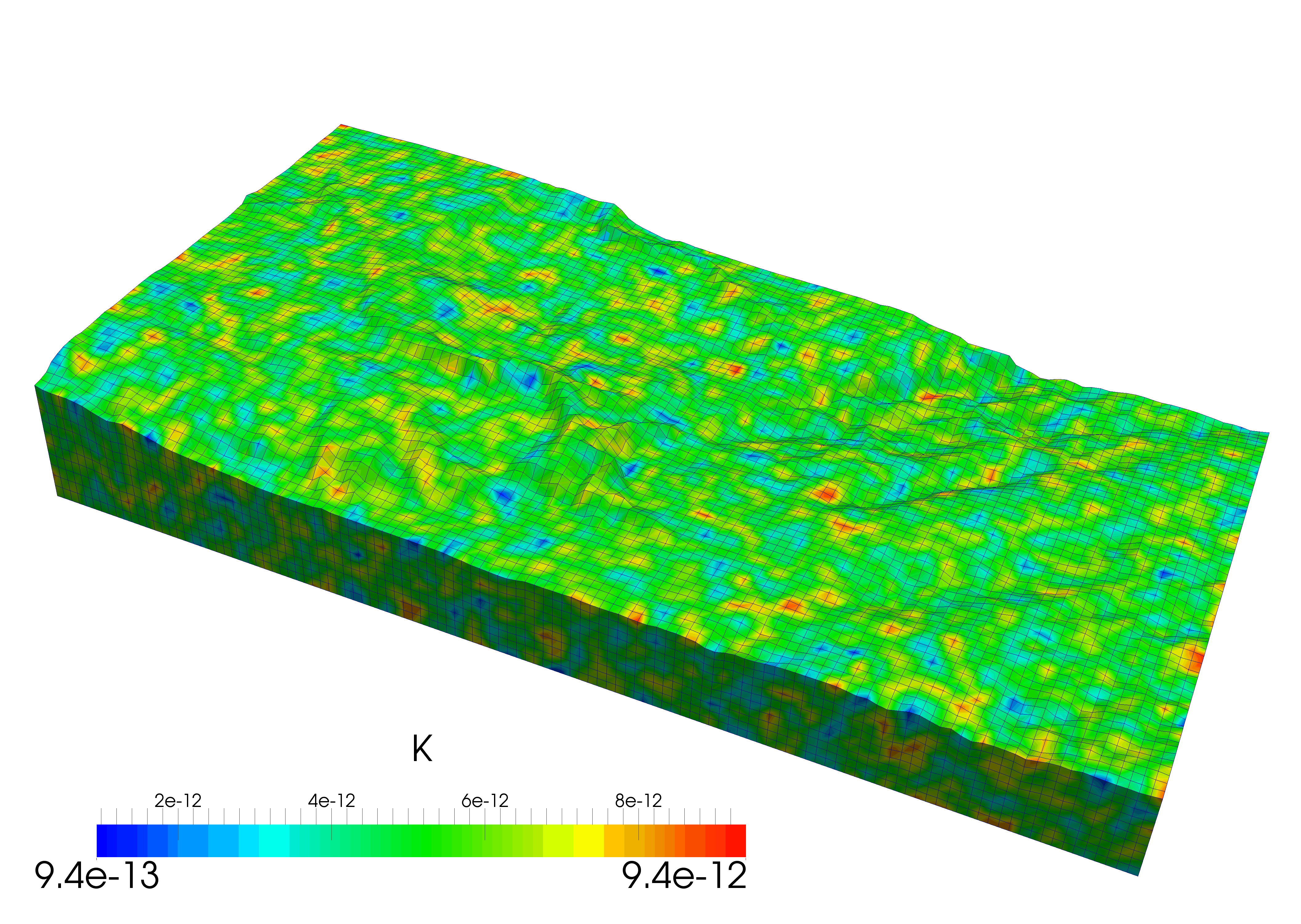}\caption{Uniformly distributed \label{fig:permeability-field}permeability
field}
\end{figure}

\begin{figure}
\centering{}\includegraphics[width=10cm]{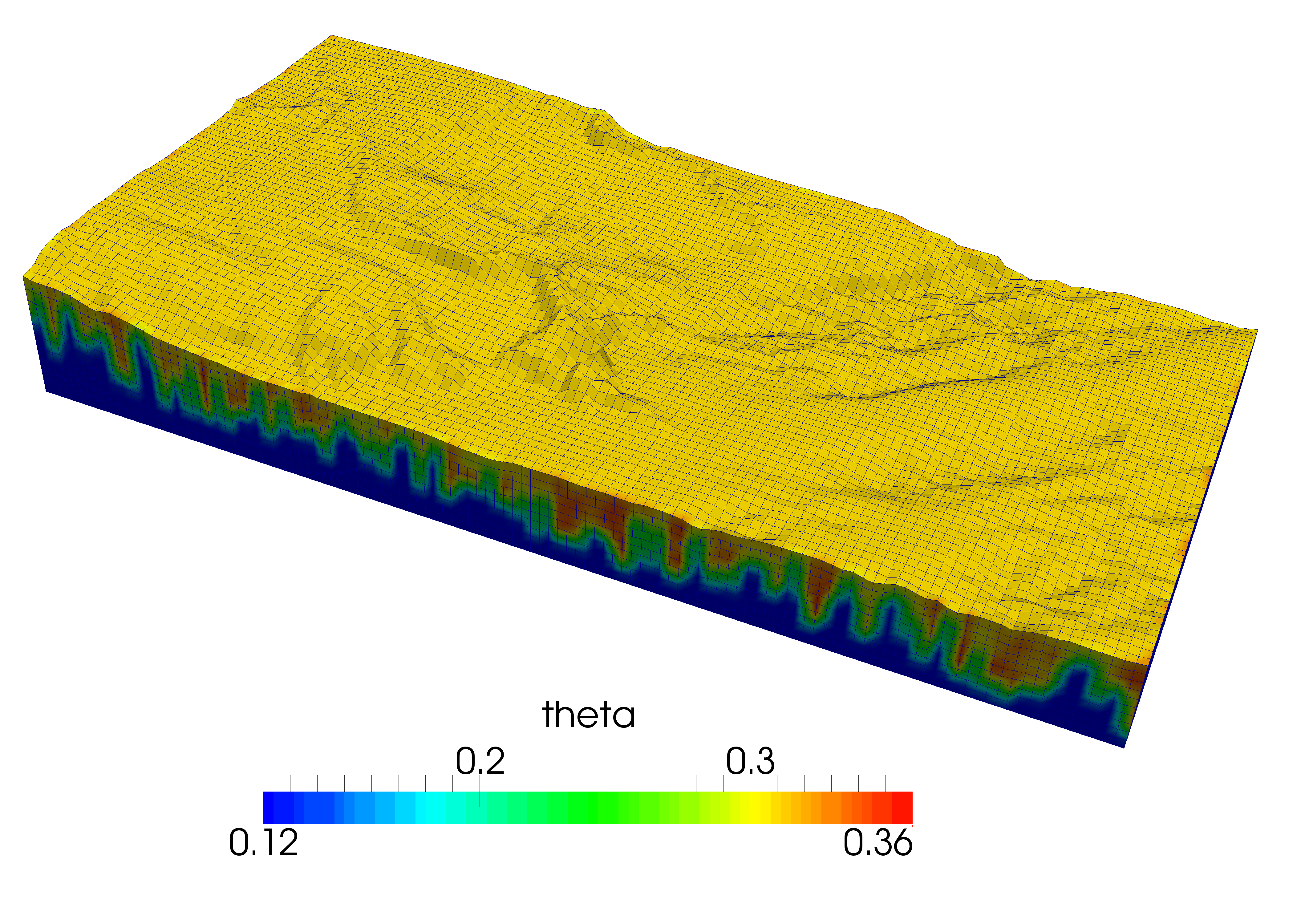}\caption{Saturation field at $t=1000$ days \label{fig:saturation-field}}
\end{figure}

The test of the parallel efficiency is performed on a 3D unstructured
mesh constructed on real topographic dataset. For this purpose, the
software MMesh3D developed by S. Marras is used \citep{Marras} which
allows to build standard mesh files in the VTK format. Using the topographic
dataset of the Monterey bay in California (dataset available with
the software), a coarse unstructured mesh composed by $60\times120\times10$
($72\,000$) computation cells is constructed in the VTK format and
then transformed into the OpenFOAM format using the utility \emph{vtkUnstructuredToFoam}.
Figure \ref{fig:Unstructured-coarse-mesh} shows the mesh with an
aspect ratio of $1:1:4$. The permeability field, randomly distributed
with a uniform law ($K\in\left[9.4.10^{-13}:9.4.10^{-12}\right]$
m\textsuperscript{2}), is shown in Fig. \ref{fig:permeability-field}.
The pressure head is initialized in the full domain with an homogeneous
value $h_{init}=-5$ m ($\theta_{init}\approx0.118$) and a fixed
pressure head $h_{top}=-0.5$ m ($\theta_{top}\approx0.306$) is imposed
on the top of the domain (the irregular face). The other parameters
used for this test are identical to those used in the Sec. \ref{sub:Validation-case}.
An example of the saturation field at $t=1000$ days using the coarse
mesh is presented in Figure \ref{fig:saturation-field}.


\begin{figure}
\centering{}\includegraphics[angle=-90,origin=c,scale=0.5]{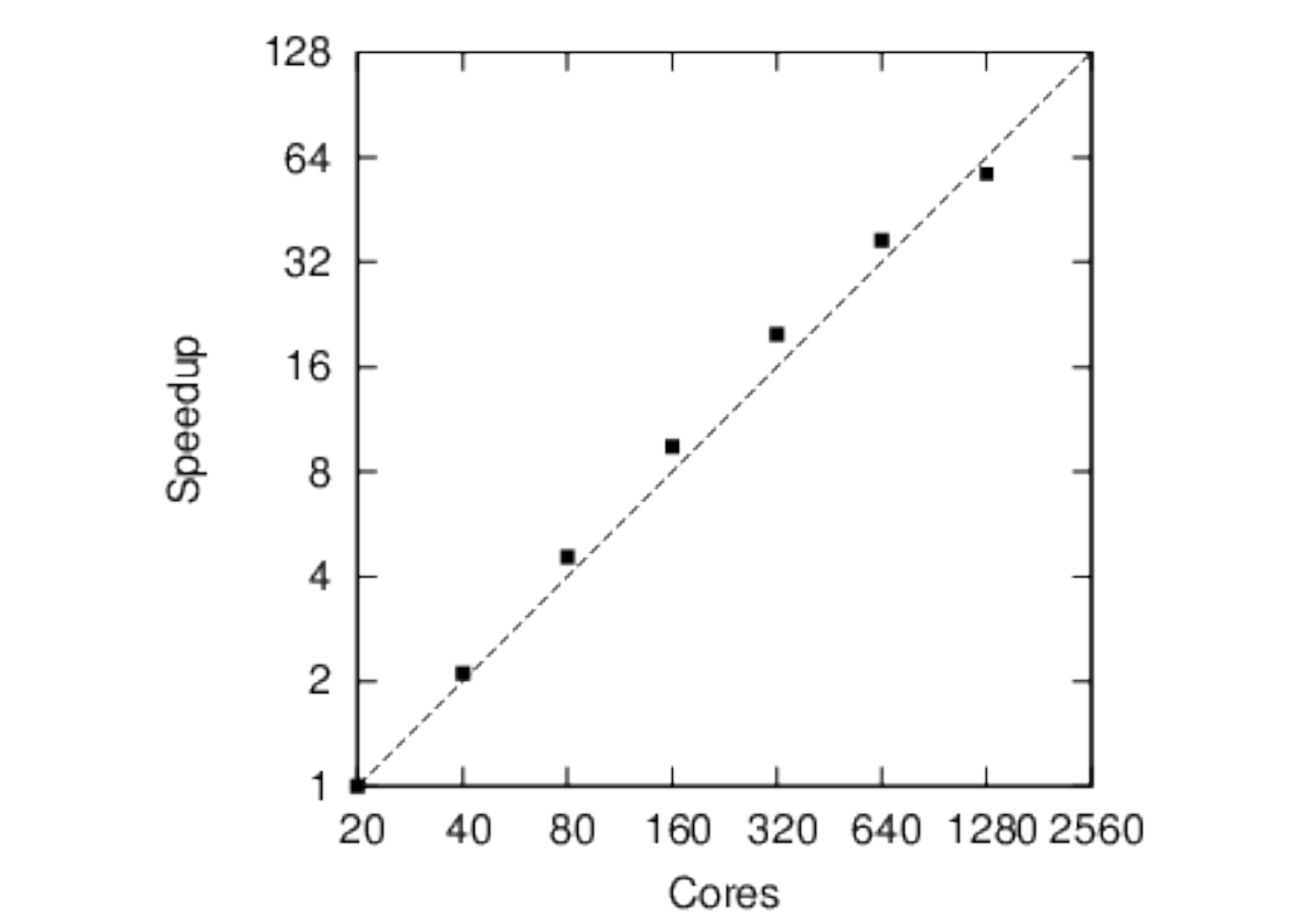}\caption{\label{fig:scalability}Log-log representation of the speedup with
the $groundwaterFoam$ solver (reference is one computation node of
$20$ cores) }
\end{figure}

To increase the size of the problem (necessary for the strong scaling
evaluation), the utility \emph{refineMesh} is used twice to multiply
by $64$ the mesh size ($240\times480\times40=4\,608\,000$ computation
cells). The infiltration phenomenon is then simulated on the CALMIP's
EOS cluster which consists of $612$ computation nodes of 2 Intel
processors 10-cores clocked at $2.8$ GHz. Simulations are performed
from $20$ (the reference) to $1280$ cores (corresponding to $64$
computation nodes) and the total CPU time required for the full simulation
is about $12$ hours. The maximum amount of memory used by the process
is $\sim5500$ Mb. The speedup $\sigma$ for a simulation with $n$
cores is computed as
\begin{equation}
\sigma_{n}=\frac{T_{20}}{T_{n}}
\end{equation}

\noindent where $T_{n}$ is the computation time for $n$ cores. The
speedup of the \textit{groundwaterFoam} solver is shown in Figure
\ref{fig:scalability} and exhibits a super-linear speedup until $640$
cores. This behavior has previously been observed with the previous
developed solver of the toolbox \citep{Horgue2015}. We should note
that the parallel efficiency is almost linear for $1280$ cores and
probably decreases for a larger number of processors. This may be
explained by the fact that the linear system for each computation
core becomes too small ($3600$ mesh cells per core for $1280$ cores).
In this configuration, the parallel efficiency allows to reduce the
computation time from $\sim34$ min ($20$ cores) to $\sim36$ seconds
($1280$ cores).

\section{Conclusion\label{sec:Conclusion}}

In this work, an OpenFOAM\textregistered{} solver dedicated to the
Richards' equation has been developed to extend the scope of the \noun{porousMultiphaseFoam}
toolbox \citep{Horgue}. The specific form of Van Genuchten's model
has been implemented to allow groundwater flow simulations with the
\emph{groundwaterFoam} solver. Three test cases are provided with
the freely accessible toolbox:
\begin{enumerate}
\item The 1D infiltration case which validates the numerical implementation
of the model by a comparison with results from the literature.
\item A 1D infiltration case with inlet velocity fixed which shows an example
of using the boundary condition \textsc{darcyGradPressure}.
\item A real topographic case with an unstructured mesh that has been used
to evaluate the parallel efficiency of the solver and exhibits a super-linear
behavior.
\end{enumerate}

\section*{Acknowledgments}

This work was granted access to the HPC resources of CALMIP under
the allocation 2013-P13147.

\section*{\noindent References}
\begin{quotation}
\noindent \bibliographystyle{plainnat}
\addcontentsline{toc}{section}{\refname}\bibliography{biblio}
\end{quotation}

\end{document}